# Surface charge writing and non-volatile control of superconductivity in LaAlO$_3$/KTaO$_3$(111) heterostructure

Siyuan Hong[1], Yanqiu Sun[1], Yuan Liu[1], Yishuai Wang[1], and Yanwu Xie[1,2]*

[1]Interdisciplinary Center for Quantum Information, State Key Laboratory of Modern Optical Instrumentation, and Zhejiang Province Key Laboratory of Quantum Technology and Device, Department of Physics, Zhejiang University, Hangzhou 310027, China

[2]Collaborative Innovation Center of Advanced Microstructures, Nanjing University, Nanjing 210093, China

*To whom correspondence should be addressed. E-mail: ywxie@zju.edu.cn

**Abstract**

The oxide interface between LaAlO$_3$ and KTaO$_3$(111) can host an electron gas that condenses into superconductivity at low temperatures. In this work, we demonstrate a local and non-volatile control of this electron gas using a biased conducting atomic force microscope tip. By scanning the tip, charges can be accumulated on the surface of LaAlO$_3$, which subsequently tune the conduction of the buried LaAlO$_3$/KTaO$_3$(111) interface largely, varying from conducting (superconducting) to insulating states. The tuning effects are stable for > 20 h at room temperature. The maximum modulation of carrier density is > $8 \times 10^{13}$/cm$^2$. This result suggests a new model system in which rewritable superconducting, normal, and insulating states can be flexibly defined in the same material on demand.

Controlling conduction locally and non-volatilely is appealing for both science and technology. Oxide heterostructures provide a fertile ground for such a control. A straightforward way to achieve it is exploiting the non-volatile polarization of epitaxial ferroelectric oxides [1,2]. By forming conducting oxide-ferroelectric heterostructures, and using electrically biased conducting atomic force microscope (cAFM) tip to locally switch the polarization, control of conduction, as well as superconductivity, has been demonstrated in a few systems including Pb(Zr,Ti)$O_3$/SrRu$O_3$ [3], Pb(Zr,Ti)$O_3$/Nb-doped SrTi$O_3$ [4], and BiFe$O_3$/YBa$_2$Cu$_3$O$_{7-\delta}$ heterostructures [5]. Even more strikingly, such a control can be achieved in an oxide heterostructure in which no ferroelectric oxide is involved at all, as discovered in LaAl$O_3$/SrTi$O_3$ (LAO/STO) [6–10]. It was proposed that LAO, as an excellent insulator, can stabilize the cAFM tip-induced charges on its surface, which in turn tunes the conduction at the buried interface [7]. Using LAO/STO as a magic template, fruitful physical phenomena and novel devices such as erasable nanoscale conduction [6], on-demand nanoelectronics [11], rewritable nanoscale oxide photodetector [12], sketched oxide single-electron transistor [13], electron pairing without superconductivity [14], one-dimensional Pascal conductance series [15], and one-dimensional Kronig-Penney superlattices [16] have been demonstrated.

Very recently, a new family of oxide interface superconductors were discovered in heterostructures composed of KTa$O_3$ (KTO) and other oxides (EuO or LAO) [17–22]. Their optimal superconducting transition temperature $T_c$ is nearly one order of magnitude higher than that of LAO/STO, and exhibits an unusual strong dependence on the orientation of KTO [17,19]. Among them, LAO/KTO resembles with LAO/STO in many aspects. Analogous to LAO/STO, the conduction in LAO/KTO locates at a few-nanometer-thick KTO layer close to the interface and is capped by a thin and highly insulating LAO film [17–19]. This key feature provides us an opportunity to control its interface conduction in a similar fashion as that in LAO/STO. In this work, using the (111)-oriented LAO/KTO heterostructure as an example, we demonstrate that we are able to write robust surface charges on LAO/KTO, and reversibly tune its interfacial conduction in a local and nonvolatile manner. The tuning effect is enormous. By varying the sign and magnitude of the tip bias, we can tune the interface into superconducting, insulating (beyond the measurement limit), and intermediate states. This control ability of superconductivity is much larger than that observed in previous ferroelectric heterostructures [4,5,23].

**Charge writing on the surface of LAO/KTO**

We first demonstrate that we can reliably write charges, as well as switch their sign, on the surface of LAO/KTO heterostructure by scanning the surface with a biased cAFM tip. Figure 1(a) schematically shows the experimental configuration. The LAO/KTO interface was grounded and a voltage bias, $V_{tip}$, was applied on the cAFM tip. Details of sample fabrication and cAFM writing can be found in the Supplemental Material (SM) [24]. For demonstration, we wrote a 3×3 μm$^2$ square with $V_{tip}$ = -9 V on a LAO(10 nm)/KTO sample, and then wrote its central 1×1 μm$^2$ square with $V_{tip}$ = +9 V. AFM topographic images (Fig. 1(b)) showed no detectable difference in the written and unwritten areas. The dusts located on the edges of the written squares were likely due to an electrostatic attraction from the biased tip. In sharp contrast, surface potential measured by Kelvin probe force microscopy (KPFM) [24] shows clear patterns (Fig. 1(c)): the area written purely with $V_{tip}$ = -9 V is dark (low potential); the central 1×1 μm$^2$ area written first with $V_{tip}$ = -9 V and followed with $V_{tip}$ = +9 V is bright (high potential). These observations resemble with that observed previously in LAO/STO [7,9], and indicate that a negative $V_{tip}$ can accumulate negative, while a positive $V_{tip}$ can accumulate positive, charges on the surface of LAO/KTO. The KPFM patterns, even the border, were robust for many hours (Figs. 1(d) and 1(e)), showing that the written surface charges are very stable.

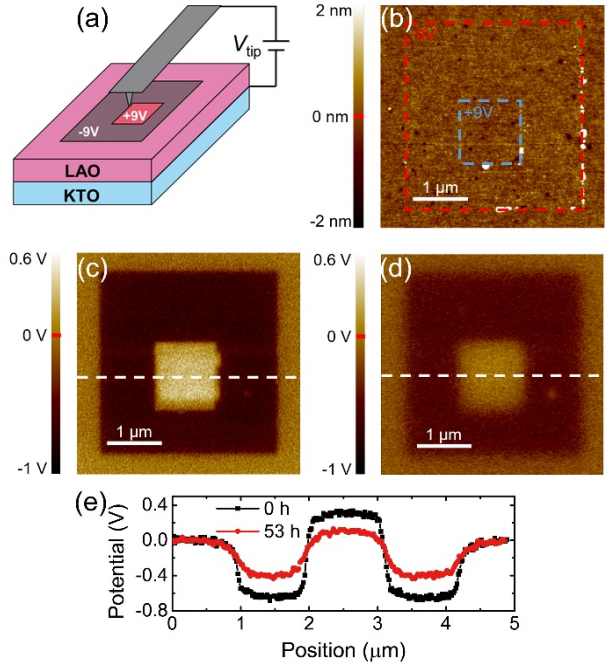

FIG. 1. Surface charge writing on a LAO(10 nm)/KTO sample. (a) Schematic for cAFM writing. The sign of $V_{tip}$ is relative to the interface. (b) AFM topographic image captured on the area immediately after the writings. The dashed squares indicate the written regions. The values of $V_{tip}$ are as labeled. (c) KPFM surface potential image measured simultaneously with the topographic image shown in (b). (d) KPFM surface potential image measured after the written sample was stored in ambient condition for 53 hours. (e) Line profiles across images (c, d) along the dashed lines.

**Mechanism for charge writing**

We then explore the mechanism for the surface charge writing. In Fig. 2 we present a KPFM surface potential image on the LAO(10 nm)/KTO surface of which we have used $V_{tip}$ = ± 9 V to write side-by-side two rectangle areas in air and another two rectangle areas under a gentle $N_2$ flow (see Fig. S1 for the experimental setup). The role of the $N_2$ flow is to create a local environment where the air that composes of ambient water vapor is replaced with dry $N_2$. As shown in Fig. 2, in contrast to the writings in air, writings under a $N_2$ flow cannot accumulate charges on the sample surface. This provides a clear evidence that ambient water vapor is crucial for the present surface charge writing. Thus we propose that the "water-cycle" mechanism found in LAO/STO [8] applies to LAO/KTO as well. In this mechanism [8], the biased cAFM tip condenses ambient water to form a water bridge between the tip and the sample surface, and at the same time dissociates some water molecular into OH$^-$ and H$^+$ (or H$_3$O$^+$) adsorbates. During the writing process, the positively (negatively) biased tip will remove some of the OH$^-$ (H$^+$ or H$_3$O$^+$) adsorbates, leaving the surface with excess positive (negative) charges [8].

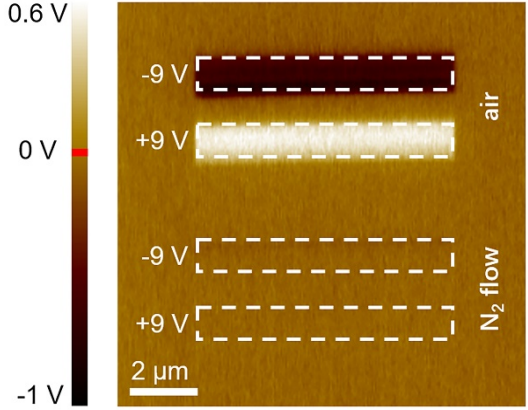

FIG. 2. KPFM surface potential image of a LAO(10 nm)/KTO sample that has been written with $V_{tip} = \pm 9$ V in both air and N$_2$ flow. The dashed rectangles indicate the written areas.

**Control the interface conduction of LAO/KTO at room temperature**

From a simple electrostatic consideration, to satisfy charge neutrality, the accumulation of negative (positive) surface charges will deplete (accumulate) interface electrons, which will in turn decrease (increase) the interface conduction. In the following we show that the charges written on the surface of LAO/KTO can indeed modulate the interface conduction hugely. In order to enable more quantitative measurements of sheet resistance $R_{sheet}$ and sheet carrier density $n_{sheet}$, we fabricated a LAO(10 nm)/KTO sample into a standard Hall-bar configuration [24] (see a photograph of the device in Fig. S2). As schemed in the inset of Fig. 3(a), we wrote surface charges on the central area containing the active conduction bridge, and then measured the transport properties using the electrodes as indicated. We found that the tuning effect of the conduction is non-volatile (with no $V_{tip}$ applied during measurements) and depends on the sign and magnitude of the applied $V_{tip}$. Roughly speaking, it is relatively easier to tune the interface to a more insulating state by applying a negative $V_{tip}$ rather than to tune the interface to a more conducting state by applying a positive $V_{tip}$. A $V_{tip} = -12$ V writing can fully suppress the room-temperature conduction to an unmeasurably tiny value, and a following $V_{tip} \geq +13$ V writing was typically needed to recover the conduction to a measurable value.

The tuned conduction is very stable. We have prepared the device into a high and low resistance states and found that both of them decayed only slightly, in ambient condition, in a time of ~20 hours (Fig. 3(a)). The tuning is reversible. By writing using $V_{tip}$ alternately with -13 V and +15 V, we could reversibly switch $R_{sheet}$ between an unmeasurably large value >2.8×10$^7$ Ω/sq and ~2×10$^4$ Ω/sq (Fig. 3(b)). These non-volatile, stable, and reversible tunings of interface conduction are in excellent agreement with the behaviors of surface charges as characterized by KPFM (Fig. 1), confirming their close correlations. Furthermore, given the fact that the $n_{sheet}$ of the unwritten LAO/KTO samples is ~5-10×10$^{13}$ cm$^{-2}$ [18,21], the ability to fully exhaust $n_{sheet}$ (become highly insulating) revealing that the tuning ability of surface charges is indeed huge.

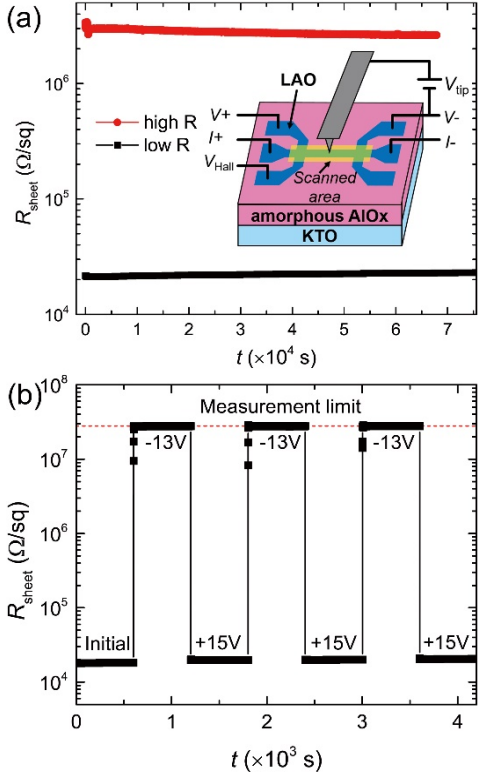

FIG. 3. Control the conduction of a LAO(10 nm)/KTO Hall-bar device. (a) Sheet resistance $R_{sheet}$ as a function of time for the device in a high-resistance state (high R) and in a low-resistance state (low R). Both states were stable and decayed only slightly after ~20 hours. Here the "high R" was prepared by scanning the device with $V_{tip}$ = -12 V first, which set $R_{sheet}$ to an unmeasurably large value, and then with $V_{tip}$ = +13 V to recover some conduction; the "low R" was prepared by scanning the device with $V_{tip}$ = +15 V. The inset shows a schematic view of writings and measurements of the Hall-bar device. The amorphous AlO$_x$ was used as a hard mask and the active LAO/KTO was confined within the Hall-bar area. Note that all the measurements were performed after (not during) writings. (b) By applying $V_{tip}$ alternately with -13 V and +15 V, we could reversibly switch $R_{sheet}$ between an unmeasurable value >2.8×10$^7$ Ω/sq and ~2×10$^4$ Ω/sq. The dashed line indicates the measurement limit. In both (a) and (b) the measurements were performed in air, at room temperature, and in four-point configurations.

**Control superconductivity**

Finally, we demonstrate our control of superconductivity of LAO/KTO using surface charge writing. It has been a long-standing challenge to control superconductivity electrostatically because the $n_{sheet}$ of most superconducting systems is larger than the tuning ability of the conventional electrical field effect [25]. Up to now, there are only handful systems, such as LAO/STO [26] and magic-angle graphene superlattices [27], which can be controlled in a sense like a conventional field effect transistor. For LAO/KTO, in a previous work [18] we have demonstrated a successful control of its superconductivity by applying an external electric field. Surprisingly, in that control, the interplay of disorders and carriers, rather than $n_{sheet}$, played a crucial role [18]. Here, the huge tuning ability resulting from the cAFM tip-induced surface charges provides us a new way to control the superconductivity of LAO/KTO.

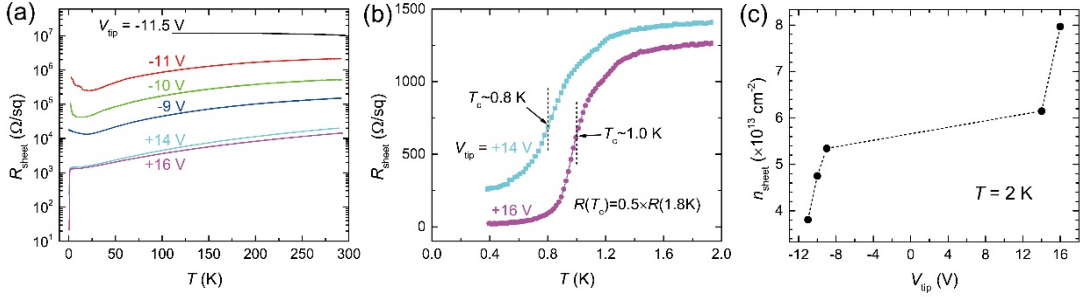

FIG. 4. Tunable transport properties in low temperatures. (a) The $R_{sheet}$ as a function of temperature for the LAO/KTO Hall-bar device scanned with different $V_{tip}$s. Before tuning, the device was in "high R" (see Fig. 3(a)). The data were measured one-by-one after subsequently scanning the device with $V_{tip}$ = +14 V, -9 V, +16 V, -10 V, -11 V, and -11.5 V. A $R_{sheet}(T)$ curve of a typical unwritten LAO/KTO sample is shown in Fig. S3 for comparison. (b) A close view of the two $R_{sheet}(T)$ curves that exhibit superconductivity. (c) The sheet carrier density $n_{sheet}$ measured at 2 K for different $V_{tip}$s.

In Fig. 4 we studied the low-temperature transport properties of the LAO/KTO Hall-bar device after it experienced charge writings with different $V_{tip}$s. To better demonstrate the creation and suppression of superconductivity, at the very beginning we set the device to a high-resistance state (see Fig. 3(a)), in which the device was too insulating to be measured in low temperatures. A following $V_{tip}$ = +14 V writing changed the device into a low-resistance state of a room-temperature $R_{sheet}$ ~$2×10^4$ Ω/sq. This time the device exhibited a metallic behavior down to ~2 K, where a superconducting transition occurred (Fig. 4(a)). The mid-point $T_c$ was ~0.8 K (Fig. 4(b)). A following $V_{tip}$ = -9 V writing fully suppressed the superconductivity and increased the overall $R_{sheet}(T)$ by about one order of magnitude (Fig. 4(a)). A further following $V_{tip}$ = +16 V writing recovered the device into a superconducting state again, with a slightly enhanced $T_c$ ~1.0 K (Figs. 4(a) and 4(b)). The subsequent writings with $V_{tip}$ = -10 V, -11 V and -11.5 V suppressed the superconductivity and increased $R_{sheet}$ gradually until an unmeasurably large value (Fig. 4(a)). This set of experiments unambiguously demonstrated that the superconductivity, as well as the overall conduction, of LAO/KTO can be efficiently controlled by the surface charge writing using cAFM tip.

To quantitatively evaluate the tuning ability of surface charges on $n_{sheet}$, we performed Hall-effect measurements immediately after the respective $R_{sheet}(T)$ measurements, on the device at 2 K. As shown in Fig. 4(c), in the measurable range the $n_{sheet}$ varied from ~3.8 to ~$8.0×10^{13}$ cm$^{-2}$. Note that here the ~$3.8×10^{13}$ cm$^{-2}$ is not the lower bound of $n_{sheet}$. It was simply because we could not make a reliable Hall-effect measurement when the channel resistance was too large due to a low $n_{sheet}$. The real lower bound of $n_{sheet}$ should be zero when the device was in a highly insulating state. Thus, the tuning ability of $n_{sheet}$ is at least up to ~$8.0\times10^{13}$ cm$^{-2}$.

In summary, we have demonstrated for the first time a non-volatile control of conduction and superconductivity at LAO/KTO interface by locally writing surface charges. The non-volatile characteristic is reminiscent of ferroelectric field-effect devices. However, unlike the ferroelectric polarization, the charge writing-induced resistive states, which depend on $V_{tip}$s, can be varied in a wide range by designing. Compared with LAO/STO [28], LAO/KTO has a higher $T_c$ and a stronger spin-orbit coupling effect [22,29–33], and in principle can work as a more friendly platform for rewritable superconducting electronic devices in which the superconducting, normal, and insulating components are realized by locally defining in the same material.

**Acknowledgements**

This work was supported by the National Natural Science Foundation of China (11934016, 12074334), the National Key Research and Development Program of China (2017YFA0303002), the Key R&D Program of Zhejiang Province, China (2020C01019, 2021C01002), and the Fundamental Research Funds for the Central Universities of China.

**Supplemental Material for**

Surface charge writing and non-volatile control of superconductivity in LaAlO$_3$/KTaO$_3$(111) heterostructure

Siyuan Hong, Yanqiu Sun, Yuan Liu, Yishuai Wang, and Yanwu Xie*

Correspondence: ywxie@zju.edu.cn

**This file includes:**

    Material and Methods
    Figs. S1-S3

**Materials and Methods.**

**Growth.** The samples were grown by depositing LaAlO$_3$(LAO) films on KTaO$_3$(KTO)(111) single crystalline substrates by pulsed laser deposition at 620 °C, in a mixed atmosphere of 1×10$^{-5}$ mbar O$_2$ and 1×10$^{-7}$ mbar H$_2$O vapor. After growth, the samples were cooled down in situ to room temperature in the same growth atmosphere. The LAO target was a single crystal. A 248-nm KrF excimer laser was used. The laser fluence was ~1 Jcm$^{-2}$. The repetition rate was 2 Hz.

**Hall bar device.** The Hall bar device was fabricated by depositing the LAO film, as described above, on a KTO(111) substrate that had been patterned using standard optical lithography and lift off techniques. To form the Hall-bar patterns, a ~200 nm-thick amorphous AlO$_x$ film was deposited as hard mask on the area outside the Hall-bar region (as schemed in the inset of Fig. 3a). The amorphous AlO$_x$ film was prepared by pulsed laser deposition at room temperature, in $P(O_2)$ = 0.01 mbar, and its interface with KTO is highly insulating (inactive). After the growth of LAO, the active LAO/KTO was confined only within the Hall-bar region. The central Hall bar bridge is 8 μm in width and 40 μm in length (see the photograph in Fig. S2).

**Atomic force microscopy (AFM), Kelvin probe force microscopy (KPFM), and charge writing.** AFM, KPFM, and charge writing were carried out using a commercial AFM machine (Park NX10) at room temperature and under ambient conditions (if not specified otherwise). Pt/Cr-coated metallic tips (Budget Sensors, Multi75E-G) with a nominal force constant of 3 N/m were used. The KPFM images were obtained in non-contact mode with a 14 nm lift height. The typical scan parameters were $V_{ac}$ of 1 V (peak-to-peak), $f_{resonace}$ of 17 kHz, scan rate of 0.2 Hz and resolution of 256×256 pixels. For charge writing, a specific $V_{tip}$ bias and a 5 nN force were applied to the conductive tips (Multi75E-G) in contact mode. During charge writing the scanning speed (the tip velocity) was of ~3.4 um/s.

**Electrical contacts and transport measurements:** The LAO/KTO interfaces were contacted by ultrasonic bonding with Al-wires. A four-probe DC method was used for all the resistance measurements. The room-temperature measurements were carried out using a Source Measure Unit instrument (Keithley 2420). The low-temperature transport measurements were performed in a commercial $^4$He cryostat with a $^3$He insert (Cryogenic Ltd.), using a current source (Keithley 6221) and a nano-voltmeter (Keithley 2182A).

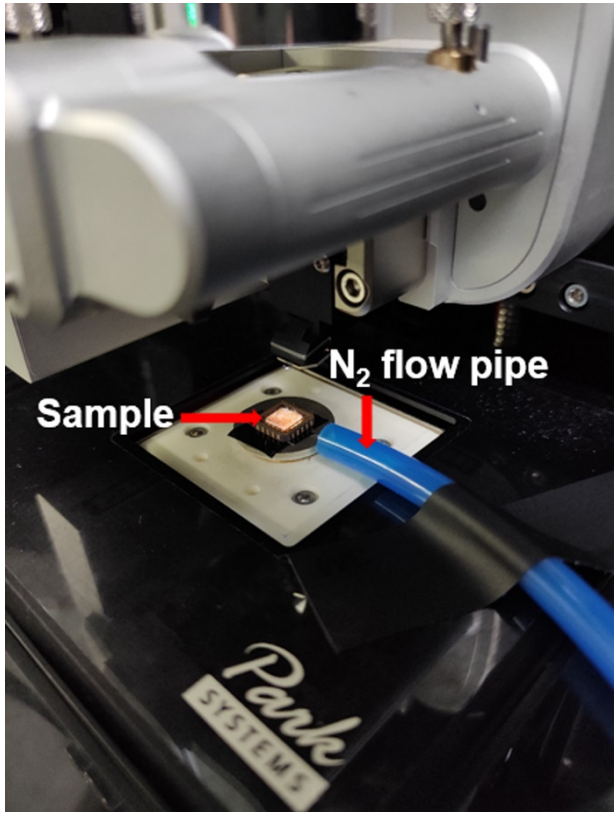

FIG. S1. Photograph of the experimental setup for writing with a biased cAFM tip on LAO/KTO under a gentle flow of dry $N_2$ gas.

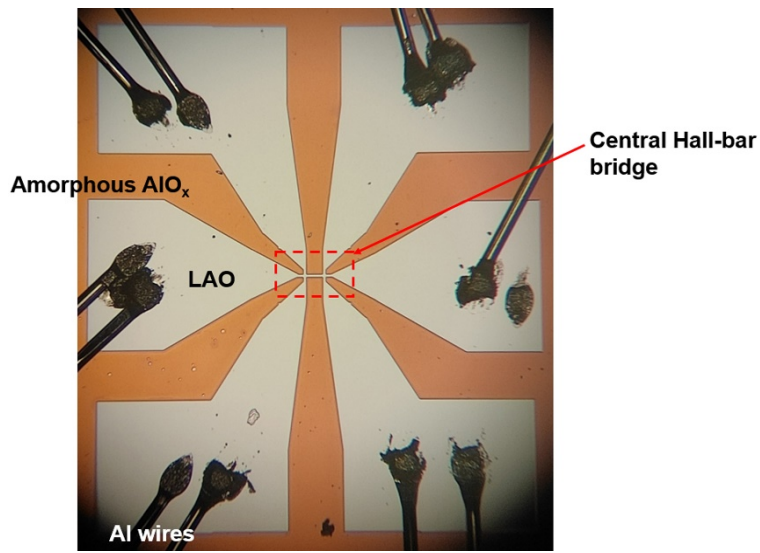

FIG. S2: Photograph of the LAO/KTO Hall device. The width and length of the central Hall-bar bridge are 8 and 40 μm, respectively.

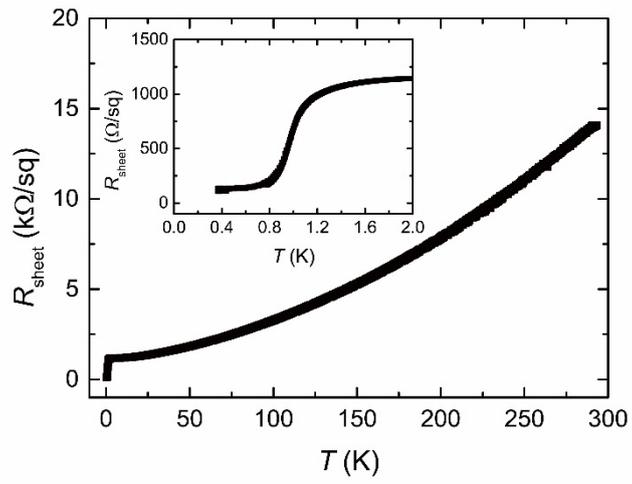

FIG. S3. Temperature-dependent $R_{sheet}$ for a typical LAO(10 nm)/KTO sample. The inset is an enlarged view of the low-temperature data.